%Paper: gr-qc/9403049
%From: BORDE@BNLCL6.BNL.GOV
%Date: Mon, 28 Mar 1994 2:11:00 -0500 (EST)

%%% Open and Closed Universes, Initial Singularities and Inflation
%%% Arvind Borde
%%% Tufts Institute of Cosmology
%%%
%%% IMPORTANT PROCESSING NOTE:
%%%       Use Plain TeX to `tex' this file, not LaTeX.
%%%       The resulting dvi file will have blank spaces in place of the
%%%       figures.
%%%       If you have a PostScript printer, *and* you use Tom Rokicki's
%%%       TeX-to-PostScript converter, DVIPS, then the figures will
%%%       also show up (after `texing' the file, type `dvips filename', then
%%%       print the PostScript file that emerges).
%%%
%%%       If you do not have a PostScript printer, or you do not use dvips,
%%%       write to borde@bnlcl6.bnl.gov and ask for the figures.
%%%
%%%%%%%%%%%%%%%%%%%%%%%%%%%%%%%%%%%%%%%%%%%%%%%%%%%%%%%%%%%%%%%%%%%%%%%%%%%%%

\newif\ifPubSub \PubSubfalse
\newcount\PubSubMag \PubSubMag=1200
\def\PubSub{\PubSubtrue
            \magnification=\PubSubMag \hoffset=0pt \voffset=0pt
            \pretolerance=600 \tolerance=1200 \vbadness=1000 \hfuzz=2 true pt
            \baselineskip=1.75\baselineskip plus.2pt minus.1pt
            \parskip=2pt plus 6pt
            \setbox\strutbox=\hbox{\vrule height .75\baselineskip
                                               depth  .25\baselineskip
                                               width 0pt}%
            \Page{5.75 true in}{8.9 true in}}

\newcount\FigNo \FigNo=0
\newbox\CapBox
\newbox\FigBox
\newtoks\0
\def\Fig {fig.~\the\FigNo}
\def\NFig {{\count255=\FigNo \advance\count255 by 1 fig.~\the\count255}}
\def\StartFigure #1#2#3{\global\advance\FigNo by 1
                  \ifPubSub \global\setbox\FigBox=\vbox\bgroup
                            \unvbox\FigBox
                            \parindent=0pt \parskip0pt
                            \eject \line{\hfil}\bigskip\bigskip
                  \else \midinsert \removelastskip \bigskip\bigskip
                  \fi
                  \begingroup \hfuzz1in
                  \dimen0=\hsize \advance\dimen0 by-2\parindent \indent
                  \vbox\bgroup\hsize=\dimen0 \parindent=0pt \parskip=0pt
                       \vrule height1pt depth0pt width0pt
                       \ifdim\dimen0<#1bp \dimen1=#1bp
                                          \advance\dimen1 by -\dimen0
                                          \divide\dimen1 by 2 \hskip-\dimen1
                       \fi
                       \hfil
                       \vbox to #2 bp\bgroup\hsize #1 bp
                            \vss\noindent\strut\special{"#3}%
                            \skip0=\parskip \advance\skip0 by\dp\strutbox
                            \vskip-\skip0 }%
\def\caption #1{\strut\egroup
                \ifPubSub \global\setbox\CapBox=\vbox{\unvbox\CapBox
                                        \parindent0pt \medskip
                                         {\bf Figure \the\FigNo:}
                                          {\tenrm\strut #1\strut}}%
                \else \bigskip {\FootCapFace \hfuzz=1pt
                                \noindent{\bf Figure~\the\FigNo:} #1\par}%
                \fi}%
\def\label (#1,#2)#3{{\offinterlineskip \parindent=0pt \parskip=0pt
                     \vskip-\parskip % To start new par.
                     \vbox to 0pt{\vss
                          \moveright #1bp\hbox to 0pt{\raise #2bp
                                         \hbox{#3}\hss}\hskip-#1bp\relax}}}%
\def\EndFigure {\egroup \endgroup % End of outmost \vbox and group.
                \ifPubSub \vfil
                          \centerline{\bf Figure \the\FigNo}
                          \egroup % End of vbox with collected figures.
                \else \bigskip \smallskip \endinsert %\StoreFigInfo
                \fi }
\def\ListCaptions {\vfil\eject \message{!Figure captions:!}%
                   \Sectionvar{Figure Captions}\par
                   \unvbox\CapBox}
\def\ShowFigures {\ifPubSub\ifnum\FigNo>0 \ListCaptions \vfil\eject
                                 \message{!Figures:!}%
                                 \nopagenumbers \unvbox\FigBox \eject
                           \fi
                  \fi}
\def\FootCapFace{} % Default

%----------------------------------------------------------------------------
%%% Layout commands:

\parskip=0pt plus 3pt
\def\Page #1#2{{\dimen0=\hsize \advance\dimen0 by-#1 \divide\dimen0 by 2
               \global\advance\hoffset by \dimen0
               \dimen0=\vsize \advance\dimen0 by-#2 \divide\dimen0 by 4
               \ifdim\dimen0<0pt \multiply\dimen0 by 3 \fi
               \global\advance\voffset by \dimen0
               \global\hsize=#1 \global\vsize=#2\relax}
               \ifdim\hsize<5.5in \tolerance=300 \pretolerance=300 \fi}
\Page{5in}{8in} % Default
\def\EndPaper{\par\dosupereject \ifnum\RefNo>1 \ShowReferences \fi
              \ShowFigures
              \par\vfill\supereject
              %???\ifPubSub
              %\else\message{Pages with (or near)
              %              PostScript figures:!\the\0}%
              %\fi
              \message{!!That's it.}\end}
\headline{\ifnum\pageno=-1 \hfil \Smallrm \Time, \Date \else \hfil \fi}
\def\VersionInfo #1{\headline{\ifnum\pageno=-1 \hfil \Smallrm #1
                              \else \hfil
                              \fi}}

%----------------------------------------------------------------------------
%%% Referencing commands:

\newcount\RefNo \RefNo=1
\newbox\RefBox
\def\Jou #1{{\it #1}}
\def\Vol #1{{\bf #1}}
\def\AddRef #1{\setbox\RefBox=\vbox{\unvbox\RefBox
                       \parindent1.75em
                       \pretolerance=1000 \hbadness=1000
                       \vskip0pt plus1pt
                       \item{\the\RefNo.}
                       \sfcode`\.=1000 \strut#1\strut}%
               \global\advance\RefNo by1 }
\def\Ref  #1{\Hunskip~[{\the\RefNo}]\AddRef{#1}}
\def\Refc #1{\Hunskip~[{\the\RefNo,$\,$}\AddRef{#1}}
\def\Refm #1{\Hunskip{\the\RefNo,$\,$}\AddRef{#1}}
\def\Refe #1{\Hunskip{\the\RefNo}]\AddRef{#1}}
\def\Refl #1{\Hunskip~[{\the\RefNo}--\AddRef{#1}}
\def\Refn #1{\Hunskip\AddRef{#1}}
\def\ShowReferences {\message{!References:!}%
                     \Sectionvar{References}\par
                     \unvbox\RefBox}
\def\StoreRef #1{\Hunskip\edef#1{\the\RefNo}}

%----------------------------------------------------------------------------
%%% Sectioning commands:

\newif\ifRomanNum
\newcount \Sno \Sno=0
\def\Interskip #1#2#3{{\removelastskip \dimen0=#1
                       \advance\dimen0 by2\baselineskip
                       \vskip0pt plus\dimen0 \penalty-300
                       \vskip0pt plus-\dimen0
                       \advance\dimen0 by -2\baselineskip
                       \vskip\dimen0 plus#2 minus#3}}
\def\Section #1\par{\Interskip{24pt}{6pt}{2pt}%
                    \global\advance\Sno by1
                    \setbox0=\hbox{\STitlefont
                                    \ifRomanNum
                                       \global\SSno=64 \uppercase
                                       \expandafter{\romannumeral
                                                    \the\Sno.\ \ }%
                                     \else
                                        \global\SSno=96 \the\Sno.\ \
                                     \fi}
                    \leftline{\vtop{\copy0 }%
                              \vtop{\advance\hsize by -\wd0
                                    \raggedright
                                    \pretolerance10000 \hbadness10000
                                    \noindent \STitlefont
                                    \GetParenDim{\dimen0}{\dimen1}%
                                    \advance\dimen0 by\dimen1
                                    \baselineskip=\dimen0 #1}}
                    \hrule height0pt depth0pt
                    \dimen0=\baselineskip \advance\dimen0 by -\parskip
                    \nobreak\vskip\dimen0 plus 3pt minus3pt \noindent}%
\def\Sectionvar #1\par{\Interskip{24pt}{6pt}{2pt}%
                    \leftline{\vbox{\noindent
                                    \STitlefont #1}}
                    \hrule height0pt depth0pt
                    \dimen0=\baselineskip \advance\dimen0 by -\parskip
                    \nobreak\vskip\dimen0 plus 3pt minus3pt \noindent}%
\newcount\SSno \SSno=0
\def\SubSection #1\par{\Interskip{15pt}{3pt}{1pt}%
                    \global\advance\SSno by1
                    \setbox0=\hbox{\SSTitlefont \char\SSno$\,$]$\,$\ }
                    \leftline{\vtop{\copy0 }%
                              \vtop{\advance\hsize by -\wd0
                                    \raggedright
                                    \pretolerance10000 \hbadness10000
                                    \noindent \SSTitlefont
                                    \GetParenDim{\dimen0}{\dimen1}%
                                    \advance\dimen0 by\dimen1
                                    \baselineskip=\dimen0 #1}}
                    \hrule height0pt depth0pt
                    \dimen0=.8\baselineskip \advance\dimen0 by -2\parskip
                    \nobreak\vskip\dimen0 plus1pt minus1pt \noindent}%
\def\(#1){~\ifRomanNum {\Medrm\uppercase\expandafter{\romannumeral #1}}%
           \else {#1}%
           \fi}

%----------------------------------------------------------------------------
%%% Equation numbering:

\newcount\EqNo \EqNo=0
\def\NumbEq {\global\advance\EqNo by 1
             \eqno(\the\EqNo)}
\def\PrevEq {(\the\EqNo)}
\def\PrevEqs #1{{\count255=\EqNo \advance\count255 by-#1\relax
                 (\the\count255)}}
\def\NameEq #1{\xdef#1{(\the\EqNo)}}

\def\AppndEq #1{\EqNo=0
  \def\NumbEq {\global\advance\EqNo by 1
             \eqno(#1\the\EqNo)}
  \def\numbeq {\global\advance\EqNo by 1
             (#1\the\EqNo)}
  \def\PrevEq {(#1\the\EqNo)}
  \def\PrevEqs ##1{{\count255=\EqNo \advance\count255 by-##1\relax
                 (#1\the\count255)}}
  \def\NameEq ##1{\xdef##1{(#1\the\EqNo)}}}

%----------------------------------------------------------------------------
%%% Theorems, etc.:

\newcount\ThmNo \ThmNo=0

\def\Theorem #1\par{\removelastskip\bigbreak
    \advance\ThmNo by 1
    \noindent{\bf Theorem \the\ThmNo:} {\sl #1\bigskip}}
\def\Lemma #1\par{\removelastskip\bigbreak
    \advance\ThmNo by 1
    \noindent{\bf Lemma \the\ThmNo:} {\sl #1\bigskip}}
\def\ContinueThm{\par\noindent}
\def\PrevThm {Theorem~\the\ThmNo}
\def\PrevLm {lemma~\the\ThmNo}
\def\Declare #1#2\par{\removelastskip\bigbreak
    \noindent{\bf #1:} {\sl #2\bigskip}}
\def\Proof {\removelastskip\bigskip
    \noindent{\bf Proof:\ \thinspace}}
\def\EndProof{{~\nobreak\hfil \copy\Tombstone
               \parfillskip=0pt \bigskip}}

%----------------------------------------------------------------------------
%%% Utilities and Miscellaneous:

\newlinechar=`\!
\def\\{\ifhmode\hfil\break\fi}
\let\thsp=\,
\def\,{\ifmmode\thsp\else,\thinspace\fi}
\def\Hunskip {\ifhmode\unskip\fi}

\def\Date {\ifcase\month\or January\or February\or March\or April\or
 May\or June\or July\or August\or September\or October\or November\or
 December\fi\ \number\day, \number\year}

\newcount\mins  \newcount\hours
\def\Time{\hours=\time \mins=\time
     \divide\hours by60 \multiply\hours by60 \advance\mins by-\hours
     \divide\hours by60         % NOTE: \divide only gives integer answers.
     \ifnum\hours=12 12:\ifnum\mins<10 0\fi\number\mins~P.M.\else
       \ifnum\hours>12 \advance\hours by-12
         \number\hours:\ifnum\mins<10 0\fi\number\mins~P.M.\else
            \ifnum\hours=0 \hours=12 \fi
         \number\hours:\ifnum\mins<10 0\fi\number\mins~A.M.\fi
     \fi }

\def\HollowBox #1#2#3{{\dimen0=#1
       \advance\dimen0 by -#3 \dimen1=\dimen0 \advance\dimen1 by -#3
        \vrule height #1 depth #3 width #3
        \hskip -#3
        \vrule height 0pt depth #3 width #2
        \llap{\vrule height #1 depth -\dimen0 width #2}%
       \hskip -#3
       \vrule height #1 depth #3 width #3}}
\newbox\Tombstone
\setbox\Tombstone=\vbox{\hbox{\HollowBox{8pt}{5pt}{.8pt}}}

\def\GetParenDim #1#2{\setbox1=\hbox{(}%
                      #1=\ht1 \advance#1 by 1pt
                      #2=\dp1 \advance#2 by 1pt}

\def\Bull #1#2\par{\removelastskip\smallbreak\textindent{$\bullet$}
                    {\it #1}#2\smallskip}
\def\Heading #1#2\par{\removelastskip\smallbreak\noindent {\it #1.}%
                      \nobreak\par #2\smallskip}

%----------------------------------------------------------------------------
%%% Title and Abstract:

\def\Title #1\par{\message{!!#1}%
                  \nopagenumbers \pageno=-1
                  \null \bigskip
                  {\leftskip=0pt plus 1fil \rightskip=\leftskip
                   \parfillskip=0pt\relax \Titlefont
                   \ifdim\baselineskip<2.5ex \baselineskip=2.5ex \fi
                   \noindent #1\par}\bigskip}
\def\Author #1\par{{\bigskip
                     \count1=0   % For address footnotes.
                     \count2=0   % For e-mail address footnotes.
                     \dimen0=0pt % For a width adjustment.
                    \Position{#1\hskip-\dimen0 }\bigskip}}
\def\Address #1\par{{\leftskip=\parindent plus 1fil \rightskip=\leftskip
                     \parfillskip=0pt\relax
                     \ifdim\baselineskip>3ex \baselineskip=3ex \fi
                     #1\par}\bigskip}
\def\modfnote #1#2{{\parindent=1.1em \leftskip=0pt \rightskip=0pt
                    \GetParenDim{\dimen1}{\dimen2}%
                    \setbox0=\hbox{\vrule height\dimen1 depth\dimen2
                                          width 0pt}%
                    \advance\dimen1 by \dimen2 \baselineskip=\dimen1
                    \vfootnote{#1}{\hangindent=\parindent \hangafter=1
                                 \unhcopy0 #2\unhbox0
                                 \vskip-\baselineskip \vskip1pt}}}%
\def\PAddress #1{\ifcase\count1 \let\symbol=\dag
                 \or \let\symbol=\ddag
                 \or \let\symbol=\P
                 \or \let\symbol=\S
                 \else \advance\count1 by -3
                       \def\symbol{\dag_{\the\count1 }}%
                 \fi
                 \advance\count1 by 1
                 \setbox0=\hbox{$^{\symbol}$}\advance\dimen0 by \wd0
                 \Hunskip \box0
                 \modfnote{$\symbol$}{{\sl Permanent address\/}: #1}}%
\def\Email #1{\ifcase\count2 \let\symbol=\ast
                 \or \let\symbol=\star
                 \or \let\symbol=\bullet
                 \or \let\symbol=\diamond
                 \or \let\symbol=\circ
                 \else \advance\count2 by -4
                       \def\symbol{\ast_{\the\count2 }}%
                 \fi
                 \advance\count2 by 1
                 \setbox0=\hbox{$^{\textstyle\symbol}$}\advance\dimen0 by \wd0
                 \Hunskip \box0
                 \modfnote{$\symbol$}{{\sl Electronic mail\/}: #1}}%
\def\Abstract{\vfil \message{!Abstract:!}%
              \Position{{\STitlefont Abstract}}
              \medskip \bgroup
              \ifdim\baselineskip>3.5ex \baselineskip=3.5ex \fi
              \narrower\noindent\ignorespaces}%
\def\EndAbstract{\par\egroup}
\def\pacs #1{\vfil\leftline{PACS numbers: #1}\eject}
\def\StartPaper{\message{!Body:!}%
                \pageno=1 \ifPubSub \footline{\tenrm\hss
                                              --\ \folio\ -- \hss}%
                          \else \footline{\hss\vtop to 0pt{\hsize=.15\hsize
                                     \vglue12pt \hrule \medskip
                                     \centerline{\tenrm\folio}\vss}\hss}%
                          \fi}

%----------------------------------------------------------------------------
%%% Typeface and other styles:

\RomanNumtrue
\let\Position=\centerline % Default positioning of certain titles.

\def\Titlefont{\tenbf}   % default
\def\STitlefont{\tenbf}  % default
\def\SSTitlefont{\tensl} % default
\def\Smallrm{\sevenrm}   % default
\def\Medrm{\tenrm}       % default

\newcount\EZReadMag \EZReadMag=1200
\def\EZRead{\magnification=\EZReadMag \hoffset=0pt \voffset=0pt
            \pretolerance=500 \tolerance=1000 \vbadness=1000 \hfuzz=1 true pt
            \baselineskip=1.1\baselineskip plus.1pt minus0pt
            \vbadness=2500
            \parskip=2pt plus 2pt
            \setbox\strutbox=\hbox{\vrule height .75\baselineskip
                                               depth  .25\baselineskip
                                               width 0pt}%
            \Page{5.75 true in}{8.5 true in}}
\EZRead

%%%%%%%%%%%%%%%%%%%%%%%%%%%%%%%%%%%%%%%%%%%%%%%%%%%%%%%%%%%%%%%%%%%%%%%%%%%%%%
\VersionInfo{March 27, 1994}

\Title OPEN AND CLOSED UNIVERSES, INITIAL SINGULARITIES AND INFLATION

\Author Arvind Borde
\PAddress{Long Island University, Southampton, NY 11968, and\\
          High Energy Theory Group, Brookhaven National Laboratory,
          Upton, NY 11973.}
\Email{borde@bnlcl6.bnl.gov}

\Address Institute of Cosmology, Department of Physics and Astronomy,\\
         Tufts University, Medford, MA 02155

\Abstract
The existence of initial singularities in expanding universes is proved
without assuming the timelike convergence condition. The assumptions
made in the proof are ones likely to hold both in open universes and in
many closed ones. (It is further argued that at least some of the
expanding closed universes that do not obey a key assumption of the
theorem will have initial singularities on other grounds.) The result
is significant for two reasons:
(a)~previous closed-universe singularity theorems have assumed
the timelike convergence condition, and
(b)~the timelike convergence condition is known to be violated in
inflationary spacetimes.
An immediate consequence of this theorem is that a recent result
on initial singularities in open, future-eternal, inflating spacetimes
may now be extended to include many closed universes.
Also, as a fringe benefit, the time-reverse of the theorem may be
applied to gravitational collapse.
\EndAbstract


\StartPaper
\Section Introduction

The singularity theorems of classical general relativity
\StoreRef{\HE}
\Refc{S.W. Hawking and G.F.R. Ellis, \Jou{The large scale structure of
      spacetime}, Cambridge University Press, Cambridge, England (1973).}
\Refe{For reviews of singularity theorems other than the one in ref.~\HE\
      see, for example, F.J.~Tipler, C.J.S.~Clarke and G.F.R.~Ellis in
      \Jou{General Relativity and Gravitation}, edited by A.~Held, Plenum,
      New York (1980).}
may be divided into two categories: those that use the
{\it timelike convergence condition\/}
($R_{ab}V^aV^b\geq0$, for all timelike vectors~$V^a$)
and those that do not. The theorems that use this condition
do so in order to make congruences of timelike geodesics
focus. Those that do not, use instead the
{\it null convergence condition\/}
($R_{ab}N^aN^b\geq0$, for all null vectors~$N^a$) in order
to make congruences of null geodesics focus. In both cases the
consequences of this focusing are then shown to be incompatible with
the other assumptions of the theorem. Theorems in the second
category include Penrose's pioneering 1965~theorem on
singularities in gravitational collapse\StoreRef{\Penrose}
\Ref{R. Penrose, \Jou{Phys. Rev. Lett.}, \Vol{14}, 57 (1965).},
Hawking's application of the time-reverse of that theorem to cosmology
\StoreRef{\HawkingO}
\Ref{S.W. Hawking, \Jou{Phys. Rev. Lett.}, \Vol{15}, 689 (1965).},
and a recent theorem on singularities in future-eternal
inflating spacetimes\StoreRef{\BVOne}
\Refc{A. Borde and A. Vilenkin, gr-qc 9312022, Tufts Institute of Cosmology
      preprint (1993).}
\StoreRef{\BVTwo}
\Refe{A. Borde and A. Vilenkin, gr-qc 9403004, to appear in the proceedings
      of the Eighth Yukawa Symposium on Relativistic Astrophysics, Japan
      (1994).}.

The timelike convergence condition implies, by continuity, the
null convergence condition, but the reverse implication is not
necessarily true: there are spacetimes~-- de Sitter spacetime is an
example~-- that violate
the timelike convergence condition but honor the
null convergence condition. The timelike convergence
condition is violated, for instance, in the inflating regions
of known inflationary spacetimes.
In fact, it has been argued~[\BVTwo] that a violation of this
condition is {\sl necessary\/} in order that a region
be considered ``inflating.'' For these reasons it is important to
prove singularity theorems without assuming the timelike
convergence condition, especially if the theorems are meant to
apply to cosmology. Such theorems exist, as mentioned above, but they
have certain weaknesses. Theorems that are directly based on
Penrose's~1965 theorem, for instance, make very strong
additional assumptions about the global structure of spacetime.
More significantly, cosmological singularity theorems
that do not assume the timelike convergence condition have all (so far)
been applicable only to open universes. Typical closed-universe
singularity theorems, on the other hand,
assume the timelike convergence condition
\Refl{R. Geroch, \Jou{Phys. Rev. Lett.}, \Vol{17}, 445 (1966).}
\StoreRef{\HawkCl}
\Refn{S.W. Hawking. \Jou{Proc. Roy. Soc. Lond.}, \Vol{A300}, 187 (1967).}
\StoreRef{\BordeCl}
\Refe{A. Borde, \Jou{Class. and Quant. Grav.}, \Vol{2}, 589 (1985).},
as do both the multipurpose 1970~theorem of Hawking and Penrose\StoreRef{\HP}
\Ref{S.W. Hawking and R. Penrose, \Jou{Proc. Roy. Soc. Lond.}, \Vol{A314},
     529 (1970).}
and Galloway's theorems extending closed-universe singularity results
\Ref{G. Galloway, \Jou{Math. Proc. Camb. Phil. Soc.}, \Vol{99}, 367
     (1986).}.
Violations of the timelike convergence condition thus provide a basis for
several apparently non-singular closed cosmologies%
\Ref{L. Parker and S.A.~Fulling, \Jou{Phys. Rev.~D}, \Vol{7}, 2357 (1973);\\
     G.L.~Murphy, \Jou{Phys. Rev.~D}, \Vol{8}, 4231 (1973);\\
     J.D.~Bekenstein, \Jou{Phys. Rev.~D}, \Vol{11}, 2072 (1975);\\
     M. Markov and V. Mukhanov, \Jou{Nuovo Cimento}, \Vol{86B}, 97 (1985);\\
     V. Mukhanov and R. Brandenberger, \Jou{Phys. Rev. Lett.}, \Vol{68},
     1969 (1992);\\
     R. Brandenberger, V. Mukhanov and A. Sornborger, \Jou{Phys. Rev.~D},
     \Vol{48}, 1629 (1993).}.

In this paper a singularity theorem is proved without assuming the
timelike convergence condition; the theorem applies
to open universes and to many closed ones. It is further
argued that some of the closed universes to which the theorem does not
apply will contain initial singularities for other reasons.
The theorem provides (among other things)
the extension to closed universes of a recent
result that demonstrates the necessity of initial singularities in
open, future-eternal, inflationary cosmologies~[\BVOne\,\BVTwo].

The paper is organized as follows:
Section\(2) discusses notation and terminology, and it gives
some background results.
Section\(3) analyzes the strategy used in some open-universe
singularity theorems.
Section\(4) discusses the recent singularity theorem that deals with
open, future-eternal, inflationary spacetimes and it shows
that this theorem, too, fits the pattern laid out in section\(3).
This analysis of open-universe theorems is important because it
suggests how one might proceed in closed universes.
Section\(5) then discusses a feature of some closed universes that
might prevent open-universe arguments from going through.
It points out that this feature does not always
occur, and it argues that the closed
universes in which it does are likely to have other
properties that force them, too, to have initial singularities.
Part of the discussion here revolves around an interesting
singularity-free spacetime due to Bardeen.
Section\(6) states and proves the main theorem of this paper and
section\(7) makes some comments on the significance of this theorem.
The paper ends with three appendices. Appendix~A briefly discusses
how the standard convergence conditions follow from conditions on the
energy-momentum tensor; it also discusses how these conditions may be
weakened from point conditions to integral ones.
Appendix~B discusses certain features of G\"odel's universe that make it
a formidable obstacle when trying to prove simple singularity theorems.
Appendix~C discusses whether the singularity predicted by cosmological
singularity theorems is indeed ``cosmological;'' i.e., whether the
theorems allow us to infer that the universe as a whole had a single
beginning.

\Section Notation and Background Results

Much of the discussion in this paper is based on the
Penrose-Hawking-Geroch ``global techniques'' in general relativity.
Everything that I need is introduced and defined below.
For further details, and for the proofs of all the assertions that
I make in this section,
see, for example, Hawking and Ellis~[\HE].

Spacetime is represented by a manifold~$\cal M$ with a Lorentz metric
$g_{ab}$ of signature $(-, +, +, +)$ defined on it. It is assumed that the
metric permits a continuous global distinction between past and future
(i.e., it is {\it time-oriented\/}). It will not be necessary
to assume a field equation in any of the arguments
given below. Convergence conditions are imposed at certain points,
and the conclusions of this paper will be valid in any theory
of gravity (such as Einstein's) in which these conditions
(or, as discussed in appendix~A, some suitable integral version)
are reasonable impositions on the curvature.

A curve in spacetime is called {\it causal\/} if it is everywhere
timelike or null (i.e., lightlike).
Let~$p$ be a point in spacetime. The
{\it causal\/} and {\it chronological pasts\/} of~$p$, denoted respectively
by $J^-(p)$ and $I^-(p)$, are defined as follows:
\smallskip

$J^-(p) = \{q:$ there is a future-directed causal curve
           from $q$ to $p\}$,

\smallskip
\noindent and
\smallskip

$I^-(p) = \{q:$ there is a future-directed timelike curve
           from $q$ to $p\}$.

\smallskip
The {\it past light cone\/} of~$p$ may then be defined%
\Ref{The expression ``light cone'' is used differently in different places
     in the literature. For instance, it is often used to refer to
     {\sl all\/} the points on null geodesics that emanate from a given
     point.
     This use of the expression is related to the definition of this paper,
     but it is not identical to it. As far as we are concerned, a point~$q$
     does not lie on the light cone of a point~$p$ if there is a timelike
     curve between~$p$ and~$q$, even if~$q$ also lies on a null geodesic
     from~$p$. (Such multiple connections cannot occur in Minkowski
     spacetime, but they can~-- and do~-- occur as soon as there is a little
     curvature.) The expression ``light cone'' is also often used in a more
     local sense, to describe just the null vectors at a point~$p$; such a
     structure will be called a ``local light cone'' in this paper.}
as~$E^-(p) = J^-(p) - I^-(p)$. It may be shown~[\HE]
that the boundaries of the two kinds of pasts of~$p$ are the same; i.e.,
$\dot J^-(p) = \dot I^-(p)$.
Further, it may be shown that $E^-(p) \subset \dot I^-(p)$.
In general, however, $E^-(p) \ne \dot I^-(p)$; i.e, the
past light cone of~$p$ (as it has been defined here) is a subset of the
boundary of the past of~$p$, but is not necessarily the full boundary
of this past. This is illustrated in \NFig. The sets~$E^-(p)$
and~$\dot I^-(p)$ (and thus also~$\dot J^-(p)$) are {\it achronal\/};
i.e., no two points on any of them can be connected by a timelike
curve. These definitions of pasts and of past light cones
may all be extended in a straightforward way
from single points to arbitrary sets.

Spacetimes in which
the type of behavior shown in \NFig\ does not occur~-- i.e., in which
$E^-(p) = \dot I^-(p)$ for all points~$p$~-- are called {\it past causally
simple}. The definition may also be tightened by further requiring that
$\dot I^-(p) \ne \emptyset$ (this rules out certain causality violations),
as was done in a previous theorem~[\BVOne\,\BVTwo]. I will not
use this tighter definition here since I will be making a separate
causality assumption.
\StartFigure{190}{115}
            {newpath 0 0 moveto
             90 90 lineto
             140 40 lineto
             170 70 lineto
             stroke
             newpath 0 0 moveto  % shaded region
             90 90 lineto
             140 40 lineto
             170 70 lineto
             190 70 lineto
             210 50 170 20 190 0 curveto
             130 -20 60 20 0 0 curveto
             gsave .9 setgray fill grestore
             2 setlinewidth
             .3 setgray
             newpath 80 70 moveto
             100 70 lineto
             stroke
             newpath 170 70 moveto
             190 70 lineto
             stroke
             .6 setlinewidth
             0 setgray
             gsave
             [3 3] 0 setdash
             newpath 160 45 moveto
             175 68 lineto
             stroke
             newpath 85 72 moveto
             90 89 lineto
             stroke
             /solidcirc { newpath
             1 0 360 arc gsave 0 setgray fill grestore} def
             90 90 solidcirc
             160 45 solidcirc
             grestore
             .3 setlinewidth
             /arrow { newpath 0 0 moveto 0 -5 lineto 2 -2 lineto
                      0 -5 moveto -2 -2 lineto stroke newpath 0 0 moveto} def
             gsave 50 58 translate arrow grestore
             gsave 115 73 translate arrow grestore
             gsave 155 63 translate arrow grestore
             newpath 10 107 moveto
             115 107 115 73 15 arcto
             115 73 lineto stroke
             newpath 10 105 moveto
             50 105 50 58 20 arcto
             50 58 lineto stroke
             newpath 167 95 moveto
             155 95 155 63 10 arcto
             155 63 lineto stroke
          }
\label(93,94){$p$}
\label(164,45){$q$}
\label(-20,104.5){E$^-(p)$}
\label(171,92){$\dot {\rm I}^-(p)-{\rm E}^-(p) \ne \emptyset$}
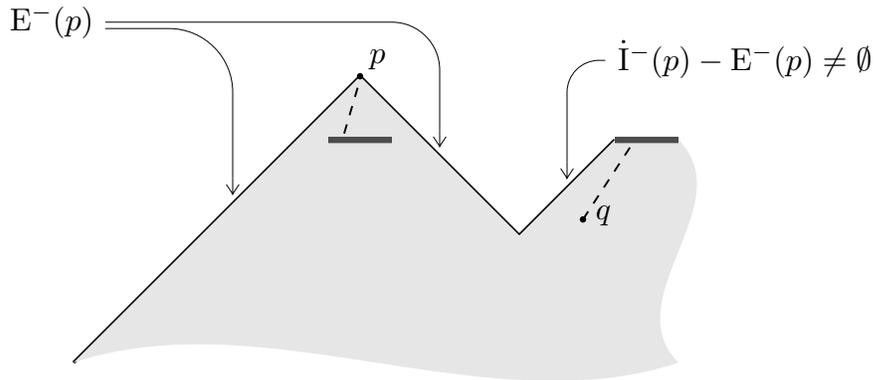
\caption{An example of the causal complications that can arise in an
unrestricted spacetime. Light rays travel along 45$^\circ$ lines in this
diagram, and the two thick horizontal lines are identified. This allows
the point~$q$ to send a signal to the point~$p$ along the dashed line,
as shown, even though~$q$ lies outside what is usually considered
the past light cone of~$p$. The boundary of the past of~$p$,
$\dot {\rm I}^-(p)$, then consists of the past light cone of~$p$,
$E^-(p)$, plus a further piece.
Such a spacetime is not ``causally simple.''}
\EndFigure

There are various kinds of causality conditions that may be
imposed, depending on which of a hierarchy of causality violations are
to be ruled out~[\HE]. The most useful of the conditions is the
{\it stable causality condition}. Roughly speaking, the condition
says that spacetime
does not contain closed timelike curves even when the metric is slightly
perturbed (i.e., the spacetime neither violates causality nor is on the
verge of doing so). A spacetime is stably causal if and only if it
admits a {\it time function\/}~[\HE]: i.e., it admits a function~$t$
whose gradient is timelike. We may assume that the gradient is
future-pointing; the function~$t$ must then strictly increase along every
future-directed timelike curve.

Another concept that we will need is that of {\it global hyperbolicity}.
Very roughly, a spacetime~$\cal M$ is globally hyperbolic if there is
a spacelike hypersurface~$\cal S$ such that the entire future and past
development of~$\cal M$ can be predicted from data on~$\cal S$.
If a surface exists with this property, it is called a
{\it global Cauchy surface\/} for~$\cal M$.
The existence of such a surface places very stringent
constraints on the global structure of~$\cal M$~[\HE].

At several points in this paper I compare closed universes
with open ones; it is useful, therefore, to define precisely
what I mean by these terms.
Intuitively a closed universe is one which ``closes on itself spatially.''
This may be made precise by saying that a closed universe is one that
contains
a closed~-- i.e., compact, without boundary~-- spacelike hypersurface.
An open universe may then be defined as one that contains no such surface.
These definitions mean that an open universe is ``open everywhere,'' but
that a closed universe is just ``closed somewhere.''
The definitions of open and closed universes may also be given a little
differently by using an {\sl achronal\/} hypersurface
instead of a spacelike one. In spacetimes without
causality violations the two definitions are closely
related. This second definition of an open universe was the one
used in the recent result on singularities
in future-eternal, inflating spacetimes~[\BVOne\,\BVTwo].

We also need some results from the theory of geodesic focusing.
Consider a congruence of causal (i.e., null or timelike)
geodesics. (A congruence is a set of curves in an open region of spacetime,
one through each point of the region.) Let~$u$ be an affine parameter
along the geodesics and let~$U^a$ be the tangent to the geodesics with
respect to this parameter. The expansion of the geodesics may be defined
as $\theta\equiv D_a U^a$, where $D_a$ is the covariant derivative.
Then the propagation
equation for $\theta$ may be written in this form~[{\HE}]:
$$
{d\theta\over du} \leq -{1\over \alpha}\theta^2 - R_{ab}U^aU^b ,\NumbEq
$$
where $\alpha=2$ for null geodesics and $\alpha=3$ for timelike geodesics.
This inequality leads to a key result on geodesic focusing:

\Lemma
Let~$\cal M$ be a spacetime in which~$R_{ab}N^aN^b\geq0$
for all null vectors~$N^a$ (i.e., the null convergence condition
holds). Consider a congruence of null geodesics with
affine parameter~$n$.
If~$\gamma$ is a member of this congruence, such that
\item{i)} the expansion~$\theta$ of the congruence is negative
on~$\gamma$ at some point~$n=n_0$, and
\item{ii)} $\gamma$ is complete in the direction of increasing~$n$ (i.e.,
$\gamma$ is defined for all~$n \geq n_0$),
\ContinueThm
then $\theta \to -\infty$ along~$\gamma$ a finite affine parameter
distance from~$n_0$.

\Proof
The proof is standard and is only sketched here; details may be found
in Hawking and Ellis~[\HE]. From~\PrevEq\ and the null convergence
condition it follows that
$$
{d\theta\over dn} \leq -{1\over 2}\theta^2 .
$$
The result is a consequence of this inequality.
\EndProof

\Section Open-Universe Singularity Theorems

Penrose's 1965~theorem on singularities in gravitational
collapse~[\Penrose] is the mother of all singularity theorems in
relativity. The theorem is based on
the existence of a {\it future-trapped surface\/}: a compact (without
boundary) spacelike 2-surface such that both systems of future-directed
null geodesics that emanate orthogonally from it (the ``inward''
system of light rays and the ``outward'' system) are converging
(i.e., have negative expansion~$\theta$).
A {\it marginally trapped surface\/} is defined similarly, but the
expansion~$\theta$ is just required to be non-positive here.
Although the concept of a trapped surface
was originally invented in order to characterize
a local collapsed system, it was soon realized by Hawking~[\HawkingO]
that it could fruitfully be put to use in cosmology as well. Hawking
pointed out that large enough 2-surfaces on constant time slices
(in terms of the usual time coordinate) of open Robertson-Walker
spacetimes are past-trapped, allowing Penrose's argument to be
applied here as well (in time-reversed form). de~Sitter-like
spacetimes also contain trapped surfaces\StoreRef{\PenroseBat}
\Ref{R. Penrose, in {\it Battelle Rencontres}, edited by
     C.M.~DeWitt and J.A.~Wheeler (W.A.~Bejamin, New York, 1968).};
this fact was exploited by Farhi and Guth
\Ref{E. Farhi and A.H. Guth, \Jou{Phys. Lett. B}, \Vol{183}, 149 (1987).}
in showing that it would be difficult to create inflationary universes
``in a laboratory.''

Closely related to the concept of a trapped surface is the idea of
a {\it reconverging light cone\/}: a point~$p$ is said to have
a reconverging past light cone if the expansion~$\theta$ of the
past-directed null geodesics in the light cone
becomes negative along every such geodesic
(i.e., the null geodesics start to converge along every past-directed
geodesic in~$E^-(p)$).
The concept was used in the 1970~theorem
proved by Hawking and Penrose~[\HP]. It was further argued there
(also see Hawking and Ellis~[\HE]) that observations of the
microwave background radiation allow us to infer how much this
radiation must have been scattered, and that this in turn implies
that there is sufficient matter along every line of sight from us
to make our own past light cone reconverge.

The significance of trapped surfaces and reconverging light
cones comes from this standard result:

\Lemma
Let~$\cal M$ be a spacetime in which~$R_{ab}N^aN^b\geq0$
for all null vectors~$N^a$ (i.e., the null convergence condition
holds). Suppose that $\cal M$ contains either a point with
a reconverging past light cone, or a past-trapped surface,
both represented here by~$\cal X$. If $\cal M$ is null-complete
to the past, then $E^-({\cal X})$ is compact.

\Proof
The result is standard, and so the proof is only sketched
here. Since~$\theta$ becomes (or already is) negative
along each of the null geodesics
that initially lies in~$E^-({\cal X})$, it follows from lemma~1 (and
the assumption of past null-completeness) that
$\theta \to -\infty$ within a finite affine parameter distance
on each geodesic. The divergence of~$\theta$ to~$-\infty$ is a
signal that the geodesics have focused. It is a standard result in global
general relativity that points on such null geodesics beyond
the focal point enter the interior of the past light cone
(i.e., enter $I^-({\cal X})$) and no longer lie in $E^-({\cal X})$~[\HE].
Thus each null geodesic that starts off in~$E^-({\cal X})$ leaves it
within a finite affine parameter distance. Since~$\cal X$ is compact,
this implies that~$E^-({\cal X})$ is compact as~well.
\EndProof

The existence of such a compact set does not by itself lead to a
singularity. To get a singularity from here, we need two additional
ingredients. First, it appears that some sort of causality assumption
is needed. Otherwise, it is possible for~$E^-({\cal X})$ to be empty
(and thus trivially compact) or, if it is not empty, for it
to be part of
a complicated enough boundary,~$\dot I^-({\cal X})$, to make further
analysis difficult. An interesting example along these lines is
given by G\"odel's universe\StoreRef{\Godel}
\Ref{K. G\"odel, in \Jou{Albert Einstein: Philosopher-Scientist},
edited by P.A.~Schilpp, Open Court Publishing Company, Chicago (1949);
\Jou{Rev. Mod. Phys.}, \Vol{21}, 447 (1949).}.
Appendix~B analyzes some features of this universe: the analysis
shows that the spacetime obeys the strict null convergence condition
and it contains both reconverging light cones and
marginally trapped surfaces
\StoreRef{\AbsBeg}
\Refc{A. Borde, Syracuse University report (1987).}
\StoreRef{\TBE}
\Refe{A. Borde, Example~V in \Jou{\TeX\ by Example}, Academic Press,
     Cambridge, MA, USA (1992).}.
Yet it is non-singular~[\HE]. This escape from a singularity occurs
because there are bad causality violations in the spacetime~[\AbsBeg\,
\Refe{R.P.A.C. Newman, \Jou{Gen. Rel. and Grav.}, \Vol{21}, 981 (1989).}.

Once causality violations are excluded
\Ref{A causality assumption appears necessary only in
     theorems that use trapped surfaces or reconverging light
     cones as their starting point. There are other theorems, which start
     differently and do not need causality assumptions~[\HawkCl\,\BordeCl].
     These theorems are, however, closed-universe theorems that assume
     the strong energy condition.},
two different approaches have been
taken in the past to obtain a singularity. One approach~--
taken, for instance, in the Hawking-Penrose theorem~[\HP]~--
uses the strong energy condition.
This approach will not be discussed here.
The other approach, which typically
uses a stronger causality assumption, delivers a
singularity immediately by ruling out the possibility of a
compact topology for~$E^-({\cal X})$. For example,
theorems based on Penrose's 1965~theorem impose
a very simple causal structure on spacetime by requiring that it
possess a global Cauchy surface~$\cal S$. Such a spacetime is necessarily
causally simple. I.e., $E^-({\cal X})=\dot I^-({\cal X})$, and thus
$E^-({\cal X})$
has no edge (being the boundary of a set). The existence of
such a compact, achronal, edgeless surface is then ruled out by further
requiring that~$\cal S$ be non-compact (i.e., that the universe be open).

Thus, the structure of a singularity theorem of this type is as follows:
\smallskip

\item{a)} It is postulated that there is a point or a set, both
represented here by~$\cal X$, with properties that lead, {\sl if the
spacetime is past null-complete}, to~$E^-({\cal X})$ being compact.
The set~$E^-({\cal X})$ is also non-empty, either by fiat, or because of
the absence of
causality violations (i.e., of closed timelike curves). The absence
of causality violations may in turn
follow from some other postulate (such as an assumption that~$\cal M$
contains a global Cauchy surface).

\item{b)} The hypersurface $E^-({\cal X})$ is edgeless. This follows
from causal simplicity~-- the assumption may be made directly or
it may follow from some other assumption (such as an assumption
that~$\cal M$ contains a global Cauchy surface).

\item{c)} The existence of the compact, achronal, edgeless hypersurface
$E^-({\cal X})$
is either asserted to be inconsistent with the structure of an open
universe, or is shown to be inconsistent with some other open-universe
assumption (such as an assumption that~$\cal M$
contains a {\sl non-compact\/} global Cauchy surface).
\smallskip

The conclusion drawn from this argument is that a spacetime cannot be
null-complete to the past under the conditions of the theorem.

\Section Inflation

A recent result~[\BVOne\,\BVTwo] shows that open
universes that eternally inflate to the future must contain initial
singularities.
A universe is said to eternally inflate to the future if the process
of inflation,
once started, never completely ends. Theoretical work as well as
computer calculations
\Refl{A. Vilenkin, \Jou{Phys. Rev.~D}, \Vol{27}, 2848 (1983);\\
      P.J. Steinhardt, in \Jou{The Very Early Universe}, edited by
        G. W. Gibbons, S. W. Hawking and S. T. C. Siklos (Cambridge
        University
        Press, Cambridge, England, 1983);\\
      A.A. Starobinsky, in \Jou{Field Theory, Quantum Gravity and Strings},
        Proceedings of the Seminar series, Meudon and Paris, France,
        1984--1985, edited by M. J. de Vega and N. Sanchez, Lecture Notes in
        Physics Vol. 246 (Springer-Verlag, New York, 1986);\\
      A.D. Linde, \Jou{Phys. Lett. B\/} \Vol{175}, 395 (1986).}
\StoreRef{\AryaVi}
\Refn{M. Aryal and A. Vilenkin, \Jou{Phys. Lett. B\/} \Vol{199}, 351 (1987);\\
      A.S. Goncharov, A.D. Linde and V.F. Mukhanov,
        \Jou{Int. J. Mod. Phys. A\/} \Vol{2}, 561 (1987);\\
      K. Nakao, Y. Nambu and M. Sasaki, \Jou{Prog. Theor. Phys.\/}
        \Vol{80}, 1041 (1988).}
\StoreRef{\Linde}
\Refe{A. Linde, D. Linde and A. Mezhlumian, \Jou{Phys. Rev.~D}, \Vol{49},
        1783 (1994).}
support the picture that inflation is indeed future-eternal: there is
an inflationary background in which new post-inflationary regions
(i.e., regions where inflation has ended) are
continually formed, but these regions never fill the entire universe.
The inflationary expansion is driven by the potential energy of a
scalar field $\varphi$, while the field slowly ``rolls down'' its
potential $V(\varphi)$. When $\varphi$ reaches the minimum of the
potential this vacuum energy thermalizes, and inflation is followed
in this region (called a ``thermalized region'') by
the usual radiation-dominated expansion. The evolution of the field
$\varphi$ is influenced by quantum fluctuations, and as a result
thermalization occurs at different times in different parts of the
universe.

A cosmological model in which new ``islands of thermalization''%
\StoreRef{\Vilenkin}
\Ref{A. Vilenkin, \Jou{Phys. Rev.~D}, \Vol{46}, 2355 (1992).}
are continually formed leads to this question:
can such a model be extended in a non-singular
way into the infinite past?
Assuming that some reasonable and rather
general conditions are met, the recently proved result shows that
in open universes the answer to this is ``no'':
such models must necessarily contain initial singularities.
This is significant, for it forces us
in inflationary cosmologies, as in the standard big-bang ones,
to face the question of what, if anything, came before.

Here is the precise statement of the result:

\Theorem
A spacetime~$\cal M$ cannot be null-geodesically complete to the past
if it satisfies the following conditions:
\item{A.} It is past causally simple, with
$E^-(x)\ne\emptyset, \forall x\in {\cal M}$.
\item{B.} It is open (i.e., $\cal M$ contains no compact, achronal
hypersurfaces without edge).
\item{C.} It obeys the null convergence condition.
\item{D.} It has at least one point~$p$ such that for some
point~$q$ to the future of~$p$ the
volume of the difference of the pasts of~$q$ and~$p$ is finite.

Assumptions~A--C are conventional as far as work on singularity theorems
goes. But assumption~D is new and is inflation-specific. It has
been discussed in detail elsewhere~[\Vilenkin\,\BVTwo], but here
is a rough, short explanation: If inflation is to be future-eternal,
then at any stage there must be a point~$p$ not on the future boundary
of the inflating region; i.e., there must be a point~$q$ a given geodesic
distance to the future
of~$p$ such that~$q$ also belongs to the inflating
part of spacetime. Now, it may be shown that if a point~$r$
lies in a thermalized region, then all points in~$I^+(r)$ are also in
that thermalized region~[\BVOne].
Further, it seems plausible that
there is a zero probability for no thermalized regions to form in an
infinite spacetime volume. Then assumption~D follows.

\Proof
The full proof of this result is available elsewhere~[\BVOne\,\BVTwo],
but here is a sketch: If~$\cal M$
is null-complete to the past, then~$E^-(p)$ must be compact.
This is so, because the volume of a small wedge of the
region $I^-(q)-I^-(p)$ around a geodesic~$\gamma$ that lies
in~$E^-(p)$ throughout may be expressed as
$$
\Delta\int_0^\infty\!\! {\cal A}(v) \,dv,
$$
where $\Delta$ is a constant, $\cal A$~is the cross sectional area
of~$E^-(p)$ around~$\gamma$, and $v$~is an affine parameter along
the geodesic (chosen to increase in the past direction).
{}From assumption~D this volume must be finite.
This can happen only if~$\cal A$ decreases somewhere. But
$$
{d{\cal A}\over dv} = \theta {\cal A}.
$$
This means that $\theta$ must become negative somewhere.
{}From assumption~C and the argument of
lemma~2 it follows that~$\gamma$ must enter $I^-(p)$, and thus
leave $E^-(p)$, within a finite affine parameter
distance. Thus~$E^-(p)$ is compact. But assumption~A implies
that~$E^-(p)$ has no edge. These two statements taken together
contradict assumption~B.
\EndProof

Thus this argument, too, follows the general open-universe pattern
laid out at the end of the previous section.

\Section Closed Universes

The open-universe pattern shows that the crucial
contradiction in open-universe singularity theorems arises from the
existence of a compact, edgeless past light cone.
The reason why closed universes prove awkward for such theorems
is that it is possible for light cones in at least some closed universes
to ``wrap around'' the whole universe and thus be compact
without causing any problems. This is illustrated in \NFig.
(As a point of interest to~\PrevThm, the past-volume
difference is finite in such a spacetime.)
Such behavior occurs, for instance, in the Einstein Universe~[\HE].
\StartFigure{92}{210}
   {% Cylinder:
    gsave
    1.5 setlinewidth
    1 .25 scale
    newpath 46 92 46 0 360 arc stroke
    newpath 46 748 46 0 180 arc stroke
    newpath 46 748 46 180 0 arc gsave [2 2] 0 setdash stroke grestore
    grestore
    newpath 0 23 moveto 0 187 lineto stroke
    newpath 92 23 moveto 92 187 lineto stroke
    % Light cone:
    newpath 20 150 moveto 15 148 5 138 0 130 curveto stroke
    newpath 20 150 moveto 40 140 72 108 92 78 curveto stroke
    gsave [2 2] 0 setdash
    newpath 0 130 moveto 20 100 52 68 72 58 curveto stroke
    newpath 92 78 moveto 87 70 77 60 72 58 curveto stroke
    grestore
    % Point:
    newpath 20 150 1.5 0 360 arc fill
    newpath 72 58 1.5 0 360 arc fill
   }
\label(23,153){$p$}
\label(75,54){$q$}
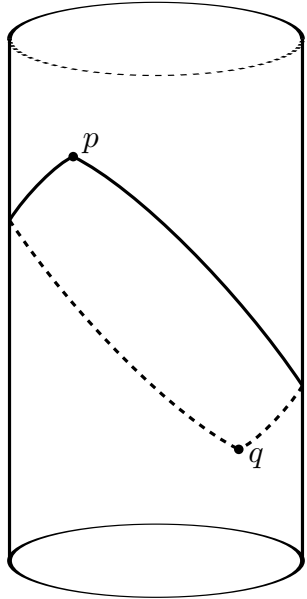
\caption{A closed universe in which the past light cone
of any point~$p$ is compact (and the volume of the difference of the
pasts of any two points is finite). The past-directed null geodesics
from~$p$ start off initially in~$E^-(p)$; but, once they recross
at~$q$ (``at the back'') they enter~$I^-(p)$ (because
there are timelike curves between~$p$ and points on these null geodesics
past~$q$), and they thus leave~$E^-(p)$.}
\EndFigure

This behavior also occurs in an interesting spacetime due to Bardeen
\StoreRef{\Bardeen}
\Ref{J. Bardeen, presented at GR5, Tiflis, U.S.S.R., and published
     in the conference proceedings in the U.S.S.R. (1968).}.
Though this spacetime was originally constructed in the
context of gravitational collapse, the lessons that it teaches are
equally relevant to the existence of initial singularities.
Bardeen's example uses the Reissner-Nordstr\"om
spacetime as inspiration.
The Reissner-Nordstr\"om metric represents the spacetime exterior
to a spherically symmetric object of mass~$m$ and electric charge~$e$.
The global properties of the spacetime depend on
the relative magnitudes of~$e$ and~$m$;
we will be interested here in the case when $e^2<m^2$.
The fully extended spacetime consists of infinitely many
regions, in each of which the standard spherical coordinates
$(t,r,\theta, \phi)$ may be used. The metric in each region is
$$
ds^2= -f(r)\,dt^2 + {1\over f(r)}\,dr^2 + r^2\,d\theta^2
      + r^2\sin^2\theta\,d\phi^2, \NumbEq
$$
where
$$
f(r)=1 - {2m\over r} + {e^2\over r^2}.
$$
This spacetime obeys the null convergence condition. Further,
there are trapped surfaces in the region $r_- < r < r_+$
(where $r_{\pm}=m \pm \sqrt{m^2 - e^2}$), and a
physical singularity at~$r=0$. But, as Bardeen pointed out~[\Bardeen],
the function~$f$ may be replaced in~\PrevEq\ by a new function~$g$,
chosen to remove the singularity~-- while still retaining
the trapped surfaces and preserving the null convergence condition.
One such function displayed by Bardeen is
$$
g(r) = 1 - {2mr^2\over (r^2 + e^2)^{3/2}},\quad r\geq0.
$$
When $e^2 < (16/27) m^2$, once again there are values~$r_{\pm}$ of~$r$
such that the region $r_- < r < r_+$ contains trapped surfaces.
The spacetime obeys the null convergence condition, yet it
contains no physical singularities.
A similar example (i.e., possessing trapped surfaces and obeying the
null convergence condition, yet non-singular) may be constructed
by directly modifying the Reissner-Nordstr\"om metric
\Ref{A. Borde, \Jou{Singularities in Classical Spacetimes}, Ph.D.
dissertation, SUNY at Stony Brook (1982).}.
This is done by choosing a value~$r_0 < r_-$ and defining the function~$g$
to agree with the Reissner-Nordstr\"om function~$f$ for $r\geq r_0$
but not for~$r< r_0$. One such function is
$$
g(r) = \cases{f(r)&\quad$(r\geq r_0)$\cr
              \noalign{\smallskip}
              \displaystyle
              1 - \left({25m^2\over 16e^2}\right)
                  \left[\Bigl({r\over r_0}\Bigr)^2 -
                  {2\over5}\Bigl({r\over r_0}\Bigr)^4\right]&\quad
                  $(0\leq r\leq r_0)$\cr}.
$$
If $r_0=4e^2/5m$ and if the parameters~$e$ and~$m$ are chosen
such that $e^2 < m^2 < (16/15) e^2$, then the new metric will have all
the desired properties.

Now, the only condition of Penrose's theorem not obeyed by
Bardeen's spacetime and its ilk (or, for that matter, by the
Reissner-Nordstr\"om spacetime) is the global Cauchy surface condition.
The lesson that is {\sl conventionally\/} drawn
from this is that the Cauchy surface assumption cannot be dropped
lightly, if we still want to prove the existence of a singularity.
If the assumption is dropped, it is usually argued,
then another strong assumption must replace
it~-- and that assumption is taken to be the timelike convergence
condition~[\HE].

\StartFigure{340}{140}
   { /PtAtInf {2 0 360 arc gsave 1 setgray fill grestore stroke} def
     /RN {% Left infinity:
          newpath
          35 0 moveto 0 35 lineto
          35 70 lineto stroke
          % Right infinity:
          newpath
          105 0 moveto 140 35 lineto
          105 70 lineto stroke
          % Trapped surface region:
          gsave
             newpath
             70 35 moveto 35 70 lineto 70 105 lineto
             105 70 lineto closepath gsave .9 setgray fill grestore
             [1 1] 0 setdash
             newpath 60 80 moveto 37 103 lineto stroke
             newpath 60 80 moveto 103 123 lineto stroke
             newpath 60 80 1.5 0 360 arc gsave 0 setgray fill grestore
          grestore
          newpath
          35 0 PtAtInf
          0 35 PtAtInf
          35 70 PtAtInf
          35 140 PtAtInf
          105 140 PtAtInf
          105 70 PtAtInf
          140 35 PtAtInf
          105 0 PtAtInf } def
       % RN spacetime:
       newpath 34 70 moveto 34 140 lineto stroke
       newpath 36 70 moveto 36 140 lineto stroke
       newpath 104 70 moveto 104 140 lineto stroke
       newpath 106 70 moveto 106 140 lineto stroke
       RN
       % Bardeen's spacetime:
       200 0 translate
       gsave
       .5 setlinewidth
       [4 4] 0 setdash
       newpath 35 70 moveto 35 140 lineto stroke
       newpath 105 70 moveto 105 140 lineto stroke
       grestore
       RN
       gsave
       1.2 setlinewidth
       newpath 35 75 moveto 55 78 85 72 105 75 curveto stroke
       grestore
   }
\label(-10,0){(a)}
\label(190,0){(b)}
\label(-23,105){$\matrix{\hfill r=0\cr \hbox{(singularity)}\cr}$}
\label(108,105){$\matrix{r=0\hfill\cr \hbox{(singularity)}\cr}$}
\label(66,78){$\cal T$}
\label(198,105){$\matrix{\hfill r=0\cr \hbox{(origin)}\cr}$}
\label(308,105){$\matrix{r=0\hfill\cr \hbox{(origin)}\cr}$}
\label(266,79){$\cal T$}
\label(280,65){$\cal S$}
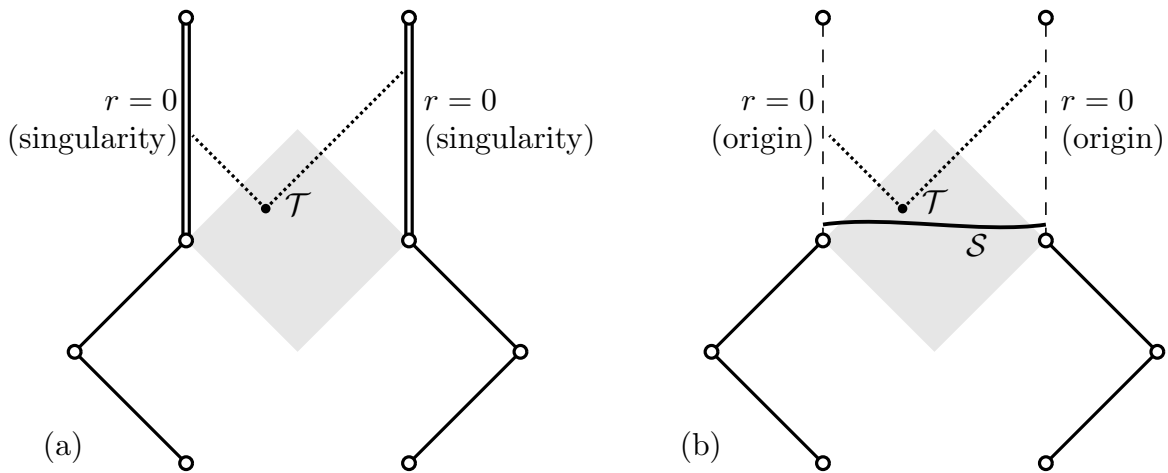
\caption{The global structure of portions of
(a)~the Reissner-Nord\-str\"om spacetime,
and (b)~Bardeen's spacetime.
A point in the interior of both spacetimes represents a 2-sphere.
The boundaries of the diagram are drawn according to these conventions:
single lines and hollow circles represent regions at infinity,
double lines represent singularities,
and dashed lines represent
the origins ($r=0$) of the coordinate systems. Thus, none
of the boundaries in~(a) are a part of the spacetime, whereas in~(b)
the $r=0$ lines are. In both cases the $r=0$ lines represent the
origins of {\sl different\/} coordinate patches.
If one imagines a series of horizontal
lines across each diagram, representing spacelike hypersurfaces, the
topology of these surfaces will be $S^2\times R$ throughout in~(a).
But in~(b) the surfaces switch from $S^2\times R$ to~$S^3$ in
the region between the $r=0$ lines. (For instance, the surface~$\cal S$
shown in~(b) is a 3-sphere.)
There are trapped surfaces,~$\cal T$, represented above by solid dots,
in the shaded regions of both spacetimes. The dotted lines emanating
from the trapped surfaces represent the two systems of future-directed
null rays from~$\cal T$: the ``ingoing'' and the ``outgoing.'' Each system
approaches~-- and in Bardeen's spacetime, reaches~-- a focal point at
$r=0$. Thus in Bardeen's spacetime (i.e., in (b)), the future light cone
of $\cal T$ spreads through the whole universe. This light cone
has topology~$S^3$.}
\EndFigure

But it is possible to draw a different lesson from Bardeen's example.
This lesson
is most clearly drawn if we compare the global structure of the
Reissner-Nordstr\"om spacetime~[\HE] with that of Bardeen's~[\Bardeen].
In the first case the topology of the spacelike sections is
$S^2\times R$ throughout. In the second it switches between~$S^2\times R$
and~$S^3$. This is illustrated in \Fig.
As the figure reveals, the escape from a singularity occurs in Bardeen's
spacetime not because it fails to be globally hyperbolic,
but because in the crucial
region where trapped surfaces occur, it is possible for light rays to
wrap around the universe. In fact, the trapped surface~$\cal T$
and its future light cone $E^+({\cal T})$ both lie
in a globally hyperbolic region of the spacetime (more precisely,
in the future Cauchy development of the surface~$\cal S$). {\sl A singularity
is avoided purely because $\cal S$ is compact}.

The wrapping of light cones around the universe (such as occurs in
Bardeen's spacetime or in the spacetime of fig.~2) ought not, however,
to be generic behavior, at least for the past cones of single points:
the cosmological scale ought to be much larger than the scale on which
light cones refocus~[\Vilenkin] (or the scale on which light rays from
``small'' trapped surfaces, such as are likely to occur in gravitational
collapse, focus). Put another way, the behavior of light cones ought to
depend only on (relatively) local effects, not on the behavior of the
universe as a whole.

There is an exception to this statement~-- one that is, unfortunately,
of interest to this paper: an expanding closed universe might well have
been small enough in the past for light rays to wrap around it easily.
Since our chief interest is the existence of initial singularities,
this scenario cannot lightly be dismissed.
But, a slight adaptation of a theorem due to Hawking~[\HawkCl]
may be used to show that in some cases an initial singularity will
exist here as well.

\Theorem
A spacetime~$\cal M$ containing a compact, edgeless, spacelike
hypersurface~$\cal S$ cannot be timelike-geodesically complete to the past
if there is a non-negative number~$K$ such that
\item{A.} $R_{ab}T^aT^b\geq-(1/3) K^2$ for all unit timelike vectors~$T^a$;
and
\item{B.} the past-directed timelike geodesics that emanate orthogonally
from~$\cal S$ have initial expansion $\theta_0<-K$ at~$\cal S$
(the past direction is the direction of increasing affine parameter).

\Proof
Suppose that~$\cal M$ is timelike-complete to the past.
Let~$\tau$ be the proper time
along the past-directed timelike geodesics from~$\cal S$.
Choose~$\tau$ to increase in the past direction and to have the value~0
at~$\cal S$. Let~$T^a$ be
the tangent to these geodesics with respect to~$\tau$ (i.e., $T^a$ is
the 4-velocity of the geodesics).
{}From formula~(1) and assumption~A we get
$$
{d\theta\over d\tau} \leq {1\over 3}\bigl(K^2 - \theta^2\bigr).
$$
This, along with assumption~B, means that
$$
\theta < K\coth \bigl({\textstyle{1\over3}}K(\tau - \hat\tau)\bigr),
$$
where $\hat\tau\equiv -(3/K)\coth^{-1}\!(\theta_0/K)>0$ (because
$\theta_0<-K<0$).
Thus $\theta\to-\infty$ within a proper time $\hat\tau$ to the past
of~$\cal S$.

Once the existence of these focal points is established, the rest of the
argument is identical to the one given by Hawking~[\HE\,\HawkCl].
\EndProof

In the original statement and proof of Hawking's result, $K$~is zero
(the focusing is then shown slightly differently than is done here).
This makes assumption~A the standard timelike convergence condition,
and assumption~B a statement that the universe is contracting in the past
direction (or, equivalently, expanding in the future direction).
The slightly different formulation given here is meant
to apply to situations where the timelike convergence condition might
not hold. Many inflationary models, for example, assume the form
$R_{ab} = 3\nu^2 g_{ab}$ for the Ricci tensor. This form satisfies
assumption~A, with $K=3\nu$
\StoreRef{\Borde}
\Ref{A. Borde, \Jou{Cl. and Quant. Gravity\/} \Vol{4}, 343 (1987).}.
Thus Hawking's theorem adapted to such situations says that there will
be an initial singularity, {\sl provided that the surface~$\cal S$ is
expanding sufficiently fast in the future direction}.
This theorem will cover
at least some cases where past light cones wrap around
the universe.

\StartFigure{92}{210}
   {% Cylinder:
    gsave
    1 .5 scale
    newpath 46 46 46 180 0 arc stroke
    newpath 46 46 46 0 180 arc gsave [2 2] 0 setdash stroke grestore
    newpath 46 374 46 0 360 arc stroke
    grestore
    newpath 0 23 moveto 0 187 lineto stroke
    newpath 92 23 moveto 92 187 lineto stroke
    /LC {/ang exch def
         newpath
         0 -10 moveto 0 0 lineto
         0 0 10 ang 270 arc closepath gsave .5 setlinewidth stroke grestore
         newpath
         0 -14 moveto 0 0 lineto
         0 0 14 ang ang arc stroke } def
    gsave 46 30 translate 225 LC grestore
    gsave 46 58 translate 200 LC grestore
    gsave 46 97 translate 180 LC grestore
    gsave 46 117 translate 155 LC grestore
    gsave
    .6 setlinewidth
    1 .5 scale
    newpath 46 106 46 180 0 arc stroke
    newpath 46 106 46 0 180 arc gsave [2 2] 0 setdash stroke grestore
    newpath 46 240 46 180 0 arc stroke
    newpath 46 240 46 0 180 arc gsave [2 2] 0 setdash stroke grestore
    newpath 46 280 46 180 0 arc stroke
    newpath 46 280 46 0 180 arc gsave [2 2] 0 setdash stroke grestore
    grestore
    newpath 46 97 1.5 0 360 arc fill
   }
\label(-54,47){spacelike $\rightarrow$}
\label(-32,114){null $\rightarrow$}
\label(-49,134){timelike $\rightarrow$}
\label(94,114){$\leftarrow\> E^-(p)$}
\label(60,123){$\gamma$}
\label(48,89){$p$}
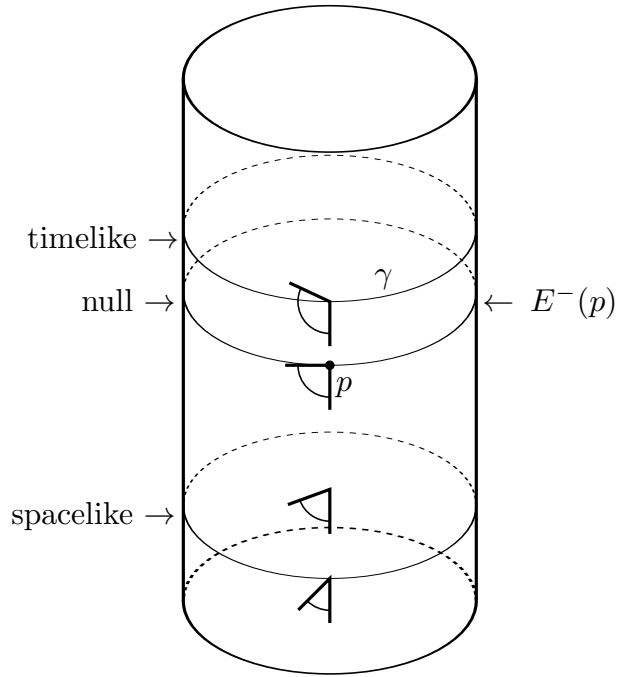
\caption{A 2-dimensional Taub-NUT-like closed universe due to
Misner~[\HE]. The causal behavior here is
very different from that in fig.~2. Local light cones are shown
at several points (only the past cones): the two straight
lines represent the two past-directed null vectors at each point, and
the arc between them the interior of the local cone.
The (global) past light cone of~$p$, $E^-(p)$,
consists of a single closed null geodesic through~$p$.
(The other null geodesic from~$p$ enters $I^-(p)$ immediately, since
past-directed timelike curves from $p$ can wind around the cylinder
and return arbitrarily close to~$p$.)
The curve~$\gamma$ to the future of~$p$ is a closed timelike curve, and
thus it does not intersect the past of~$p$. This means that even though
in a certain sense $E^-(p)$ wraps around the universe, points to the
future of this set can avoid sending
signals that intersect it no matter how close they lie to~$E^-(p)$.}
\EndFigure

For the rest of this paper I will concentrate on situations where
light cones do not wrap around the universe. A light cone such as the
one in~fig.~2 that wraps around also swallows the universe
entirely~[\PenroseBat]. One feature of this ``swallowing'' may be seen
if we examine a point close to the light cone, but to its
future: it appears impossible for {\sl any\/} past-directed signal from
the point to avoid intersecting the cone
\Ref{In fact, in fig.~2 even a point far from the light cone, but still
     to its future, will also have this property. But the statement
     here is phrased in terms of a point that is ``close'' so as to
     cover other
     situations where there might be topological interconnections or
     singularities far from the light cone that divert or block signals
     from reaching it.}.
Actually, appearances are a little deceptive here:
these statements are not true, as \Fig\ shows,
in causality-violating spacetimes of the Taub-NUT variety~[\HE].
But if we exclude causality violations, we may take the statements made
above as characterizing the behavior that we want to exclude.
I.e., I will assume in the theorem that past light cones are
localized in the following sense:

\Declare{Definition} A past light cone in a stably causal
spacetime is called {\it localized} if from every spacetime
point~$p$ not on the cone there is at least
one past-directed timelike curve that does not intersect the cone.

This idea is illustrated in \NFig.
Spacetimes in which past light cones are localized include
all the standard open universes (Minkowski, Schwarzschild, the
open Robertson-Walker cosmologies, etc.), as well as closed ones
such as de~Sitter and some of the closed Robertson-Walker cosmologies
(here, the localization or not of light cones depends on the time scale
for the recollapse of the universe).

\StartFigure{275}{100}
    {% Example (a):
     gsave
     newpath 50 0 moveto 0 50 lineto 50 100 lineto 100 50 lineto
     closepath clip
     .5 setlinewidth
     newpath 0 10 moveto 60 70 lineto 100 30 lineto
     100 0 lineto 0 0 lineto closepath
     gsave .9 setgray fill grestore stroke
     1 setlinewidth
     newpath 50 69 moveto
     [1 1] 0 setdash 40 56 7 20 0 14 curveto stroke
     newpath 50 69 1.5 0 360 arc fill
     grestore
     newpath 50 0 moveto 0 50 lineto 50 100 lineto 100 50 lineto
     closepath stroke % boundary
     % Example (b):
     gsave
     newpath 150 80 moveto 160 60 265 60 275 80 curveto
             265 100 160 100 150 80 curveto stroke
     gsave [2 2] 0 setdash .5 setlinewidth
     newpath 150 20 moveto 160 40 265 40 275 20 curveto stroke
     grestore
     gsave
     .5 setlinewidth
     newpath 150 20 moveto 150 80 lineto stroke
     newpath 275 20 moveto 275 80 lineto stroke
     grestore
     newpath 150 80 moveto 150 20 lineto 160 0 265 0 275 20 curveto
     275 80 lineto closepath clip
     .5 setlinewidth
     newpath 170 0 moveto 220 50 lineto 270 0 lineto
     gsave .9 setgray fill grestore stroke
     1 setlinewidth
     newpath 210 47 moveto
     [1 1] 0 setdash 200 34 177 10 173 7 curveto stroke
     newpath 210 47 1.5 0 360 arc fill
     grestore
     newpath 150 20 moveto 160 0 265 0 275 20 curveto stroke % bottom edge
    }
\label(40,68){$p$}
\label(200,46){$p$}
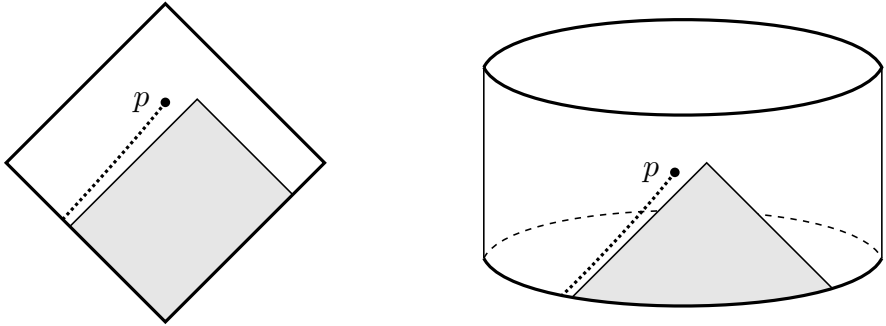
\caption{Two examples of spacetimes with localized past light cones. The
boundaries of both figures are boundaries at infinity.
A typical past light cone is shown in each spacetime.
The dotted line from each point~$p$ represents a past-directed timelike
curve that avoids intersecting the past light cone.}
\EndFigure

\Section The Singularity Theorem

If a past light cone is compact and without edge, it is not ``localized.''
In fact, a slightly more general result holds:

\Lemma
Let~$\cal M$ be a spacetime that obeys the stable causality condition.
Suppose that $\cal M$ contains a
compact, achronal hypersurface, $\cal S$, without edge.
Then there are points~$p$ in~$I^+({\cal S})$ such that every past-directed
timelike curve from~$p$ intersects~$\cal S$.

\Proof
Let $t$ be a time function on~$\cal M$. Vary~$\cal S$ forward a small amount
in the future~$t$ direction. This gives a compact spacetime region~$\cal N$
with two compact components to its boundary: $\cal S$, and a second
component
denoted by~${\cal S}'$ (see \NFig). Though~$\cal S$ is achronal by
assumption, ${\cal S}'$ does not have to be.
Let~$t_0$ be the minimum value of~$t$ on~${\cal S}'$, attained at some
point~$p$. Every past-directed timelike curve from~$p$ must
enter~$\cal N$. This is so, because~$t_0$ is the minimum value of~$t$
on~${\cal S}'$ and, further, because ${\cal S}'$ has no edge.
Each such curve~$\gamma$ must also eventually leave~$\cal N$:
if it does not, it must accumulate at some point in the compact
set~$\cal N$, and examination of constant~$t$ surfaces in a small
neighborhood of the accumulation point shows that this cannot happen
(because $t$~decreases along~$\gamma$).
The curve must leave through~$\cal S$ and not through some other
point of~${\cal S}'$ (again, because $t$~decreases along the curve).
Thus, every past directed timelike curve from~$p$ intersects~$\cal S$.
\EndProof

\StartFigure{120}{100}
   {newpath 10 20 moveto 17 0 118 -30 110 20 curveto
    110 80 lineto
    118 30 17 60 10 80 curveto
    10 20 lineto gsave .95 setgray fill grestore
    newpath 10 20 moveto 17 0 118 -30 110 20 curveto stroke
    newpath 10 80 moveto 17 60 118 30 110 80 curveto stroke
    newpath 10 80 moveto 3 100 102 130 110 80 curveto stroke
    newpath 10 20 moveto 3 40 102 70 110 20 curveto
            gsave [2 2] 0 setdash .5 setlinewidth stroke grestore
    gsave .6 setlinewidth
    newpath 10 20 moveto 10 80 lineto stroke
    newpath 110 20 moveto 110 80 lineto stroke
    grestore
    newpath 80 52.5 1.5 0 360 arc fill
    [1 1] 0 setdash
    newpath 80 52.5 moveto 78 40 73 25 65 2 curveto stroke
   }
\label(0,20){$\cal S$}
\label(0,80){${\cal S}'$}
\label(17,45){${\cal N}$}
\label(78,59){$p$}
\label(66,30){$\gamma$}
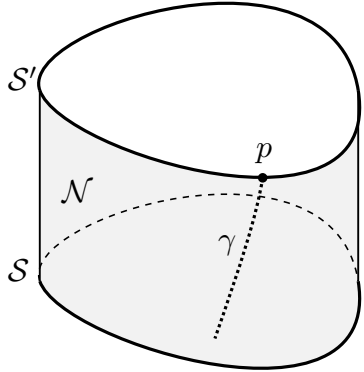
\caption{An illustration of the strategy used in~\PrevLm.}
\EndFigure

The main result of this paper follows trivially from the preceding
discussion:

\Theorem
A stably-causal spacetime cannot simultaneously satisfy
the following two conditions:
\item{A.} It is past causally simple.
\item{B.} It contains a compact, localized past light cone.

\Proof
The proof follows immediately from what has gone before.
If~$\cal E$ is the past light cone given by assumption~B, then
assumption~A implies that~$\cal E$ has no edge (being
the full boundary of the past). Also, $\cal E$ is achronal.
This contradicts lemma~5.
\EndProof

This theorem (backed by lemma~2 and Theorem~3) shows
that spacetimes cannot be
past null-complete in a variety of circumstances:

\Declare{Corollary}
A stably-causal spacetime~$\cal M$ cannot be null-geodesically complete to
the past if it satisfies the following conditions:
\item{A.} It is past causally simple.
\item{B.} All past light cones in~$\cal M$ are localized.
\item{C.} It obeys the null convergence condition.
\item{D.} It contains one of these:
\itemitem{i)} a point with a reconverging past light cone, or
\itemitem{ii)} a past-trapped surface, or
\itemitem{iii)} a point~$p$ such that for some point~$q$ to the future
of~$p$ the volume of the difference of the pasts of~$q$ and~$p$ is finite.

This result offers (among other consequences) a modification of Theorem~3
by replacing its assumption~B by the assumptions that the
spacetime is stably causal, and that all past light cones in the
spacetime are localized.
The modified theorem will not exclude closed universes, as the
original one had.

\Section Discussion

The result presented above has many of the strengths
of the other singularity theorems, and it also shares many
of their weaknesses. It is based on very modest assumptions,
and it may thus be considered a strong result.
But the conclusions that it comes to are also somewhat limited.
The theorem demonstrates the
existence of a past-incomplete null geodesic, but it yields
no information on where in the past the singularity lies.
Nor does it demonstrate that the universe had a ``single
beginning'' in the sense that Robertson-Walker models might
be said to have one. The question of a single beginning is addressed
further in appendix~C.

Still, the theorem closes several gaps in our understanding
of the conditions that are likely to lead to, or to prevent,
singularities. It shows, for instance, that non-singular cosmologies
must either violate the null convergence condition (in addition to
the violations of the timelike convergence condition that
many are already known to possess), or they must not have reconverging
past light cones, or if light cones do reconverge then they must swallow
the universe. (If there is a violation of the null
convergence condition, it must be severe enough to also violate the
integral conditions discussed in appendix~A.)
The theorem also places the
initial singularity in the past, where it rightfully belongs,
unlike (say) the Hawking-Penrose theorem~[\HP] which is entirely
silent on the location of the singularity.

Apart from its cosmological applications, the theorem may also
be applied to gravitationally collapsing systems
in which future-trapped surfaces occur. A future singularity
is predicted here by the time-reverse of the
theorem, thus covering spacetimes like
Reissner-Nordstr\"om (which, oddly, has not hitherto met
the conditions of any singularity theorem
\Ref{The multipurpose Hawking-Penrose theorem~[\HP] comes close
     to embracing the Reissner-Nordstr\"om spacetime, but
     its ``generic condition'' does not hold there.
     (Though the original impetus
     for the development of singularity theorems came from the need to
     show that singularities were not just a feature of special solutions,
     but were generic, the adoption of the ``generic condition'' in the
     Hawking-Penrose theorem has had the effect of only covering
     generic situations, and excluding special ones.)}).

The theorem here is, of course, only as good as its assumptions:
the weaker we make the assumptions, the stronger
and more physically reliable the result.
One condition that can probably be weakened somewhat is the
causal simplicity assumption.
This assumption was made solely to prove the
theorem in the most direct way possible and with a minimum of
mathematical fuss. But it would be preferable to have stable
causality as the only restriction on causal structure.
This issue will be discussed elsewhere.

It would also be nice if the
assumption on localized null cones could be replaced by something
likely to hold in all closed universes. But the existence of
non-singular closed spacetimes that obey the null convergence
condition makes it far from obvious (at least to me) how
to proceed without such an assumption~-- though it is possible
that some kind of genericity condition might help here.
In this connection, it
is curious that closed-universe singularity theorems have tended
to need stronger convergence conditions than open-universe theorems,
despite the fact that a closed universe is presumably denser than an open
one and thus
ought to have a greater natural proclivity for singularities.

\AppndEq{A}
\Sectionvar Appendix A: Convergence Conditions and Energy

Both of the standard convergence conditions, timelike and null,
follow, via Einstein's equation, from certain inequalities~--
known as energy conditions~-- on the
matter energy-momentum tensor~[\HE]. Einstein's equation is
$$
R_{ab} -{1\over 2}Rg_{ab} + \Lambda g_{ab} = 8\pi T_{ab}, \NumbEq
$$
where $R_{ab}$ is the Ricci tensor obtained from~$g_{ab}$, $R$~is the
curvature scalar, $\Lambda$ is the cosmological constant, and $T_{ab}$
is the matter energy-momentum tensor.
The inequalities that are useful for our purposes are the
{\it strong energy
condition\/} (which says that
$T_{ab}V^aV^b - (1/2)T^a_{\hphantom{a}a}V^bV_b \geq 0$ for
all timelike vectors~$V^a$) and the {\it weak energy condition\/}
(which says that $T_{ab}V^aV^b \geq 0$
for all timelike vectors~$V^a$, from which it follows by continuity
that $T_{ab}N^aN^b \geq 0$ for all null vectors~$N^a$).
The timelike convergence condition
follows from the strong energy condition when the cosmological
constant is zero, and the null convergence condition
follows from the weak energy condition, even if there is a
cosmological constant.

All these conditions are point conditions, and there has been discussion
\StoreRef{\Tipler}
\Refl{F.J. Tipler, \Jou{J. Diff. Eq.}, {\bf 30}, 165 (1978);
      \Jou{Phys. Rev.~D}, {\bf 17}, 2521 (1978).}
\StoreRef{\CE}
\Refn{C. Chicone and P. Ehrlich, \Jou{Manuscripta Math.}, \Vol{31},
      297 (1980).}
\Refn{T. Roman, \Jou{Phys. Rev.~D}, {\bf 33}, 3526 (1986);
      \Jou{Phys. Rev.~D}, {\bf 37}, 546 (1988).}
\Refn{M.S. Morris, K.S. Thorne and U. Yurtsever, \Jou{Phys.\ Rev.\ Lett.},
      \Vol{61}, 1446 (1988).}
\Refn{U. Yurtsever, \Jou{Class. Quant. Grav.}, \Vol{7}, L251 (1990).}
\Refn{G. Klinkhammer, \Jou{Phys. Rev.~D}, \Vol{43}, 2542 (1991).}
\Refe{R. Wald and U. Yurtsever, \Jou{Phys. Rev.~D}, \Vol{44}, 403 (1991).}
of scenarios in which the energy conditions
might be violated in a limited way (for example, at some points but
not at others). It is known from some of this work~[\Tipler\,\CE\,\Borde]
that what is important in order to ensure focusing is that~$R_{ab}U^aU^b$
(where~$U^a$ is the tangent to a null or a timelike geodesic)
obey an integral inequality, not necessarily one that must hold
at each point. Such
integral~-- or, as they have come to be called, averaged~-- convergence
conditions will do just as well for the purposes of this paper.
For instance, a condition that (roughly speaking) requires
that $\int\! R_{ab}N^aN^b\,dn$ be
repeatedly non-negative (along a geodesic with affine parameter~$n$)
is known to be sufficient to give focusing~[\Borde].

Despite the availability of these weaker conditions, the
central arguments of this paper are phrased in terms of point
conditions. This is done in order to keep the line of the argument clean
and uncluttered with extraneous detail. It should be kept in mind,
though, that any convergence condition~-- point or integral~--
that is sufficient to guarantee that a congruence of initially
converging, complete geodesics actually comes to a focus is adequate
for our purposes.

\Sectionvar Appendix B: G\"odel's Universe

In the course of an investigation of the idealistic conception of time
(i.e., of whether or not ``reality consists of an infinity of layers of
`now'\thinspace''), Kurt G\"odel discovered in 1949 a very interesting
solution to Einstein's equation~[\Godel]. The solution has
closed timelike curves. It is also simply connected. It follows from this
that G\"odel's universe contains no edgeless spacelike hypersurfaces:
``reality'' here does not contain even a single layer of ``now.''
Another interesting feature of this universe~-- though less dramatic~--
is that it has no singularities, despite (as we shall see below)
possessing reconverging light cones and marginally trapped
surfaces, and obeying the strict null convergence condition
\Ref{The strict condition is that $R_{ab}N^aN^b$ is strictly positive for
all null vectors~$N^a$. Under this condition, the light rays from even
a marginally trapped surface will focus.}.
This makes it
a tantalizing obstacle when trying to develop singularity theorems
\Ref{My attention was drawn to this aspect of G\"odel's solution
     by Roger Penrose.}.

The manifold on which G\"odel's metric is defined is~$\Re^4$.
A set of coordinates $(t', x, y, z)$ may be chosen such that each
coordinate has range $(-\infty,\infty)$, with the metric given by
$$
ds^2=-{dt'}^2+dx^2-{\textstyle{1\over2}}\exp{(2\sqrt2\omega x)}\,dy^2+dz^2
-2\exp{(\sqrt2\omega x)}\,dt'dy,
$$
where $\omega$ is a positive constant.
The metric satisfies Einstein's equation, with a cosmological constant
(see equation~\PrevEq),
if $T_{ab}=\varrho U_aU_b$,
where~$U^a= (\partial/\partial t')^a$, and $\omega^2=-\Lambda=4\pi\varrho$.
If~$N^a$ is a null vector, we have $R_{ab}N^aN^b=2\omega^2 (U_aN^a)^2 >0$;
i.e., the strict null convergence condition holds here.

The coordinates $(t', x, y, z)$ are not the best ones in
which to investigate light cones.
New coordinates $(t, r, \phi , z)$ may be defined~[\HE], with
$-\infty< t <\infty$, $0\leq r<\infty$, $0\leq\phi\leq 2\pi$, and
$-\infty<z<\infty$, by the following transformations:
$$
\eqalignno{\exp{(\sqrt2\omega x)} & = \cosh{2r}+\cos\phi\,\sinh{2r}, &\cr
\omega y\exp{(\sqrt2\omega x)} & = \sin\phi\,\sinh{2r}, &\cr
\tan{\textstyle{1\over2}}(\phi + \omega t - \sqrt2t') & = \exp(-2r)
\tan{\textstyle{1\over2}}\phi, &\cr
\noalign{\hbox{and}}
z&=z.&\cr}
$$
In these coordinates, the metric is given by
$$
ds^2={2\over{\omega^2}}\bigl(-dt^2+dr^2-
\sinh^2\!r\,(\sinh^2\!r-1)\,d\phi^2+2\sqrt2\sinh^2\!r\,d\phi\,dt\bigr)
+dz^2.
$$
The coordinate~$z$ plays no role in the behavior of the spacetime,
and it will be ignored from now on.
It may be checked that the closed curve given by
$$
t=\hbox{constant}, \quad r=\log(1+\sqrt2), \quad {\rm and} \quad
\phi=\phi(p)
$$
is a nongeodesic null curve~[\TBE] (where~$p$ is an arbitrary parameter.)
But, in the region $r < \log(1+\sqrt2)$, the surfaces of constant~$t$
and~$r$ are spacelike. (They are also compact and edgeless~--
if we ignore~$z$, or if we compactify the~$z$ direction by identifying,
say, $z=0$ with~$z=1$.)
The tangents to the two systems
of past-directed null geodesics that emanate orthogonally from these
surfaces have these non-zero components:
$$
\eqalign{N^t_{\pm}&={\sinh^2\!r-1\over\cosh^2\!r},\cr
N^r_{\pm}&=\pm{(1-\sinh^2\!r)^{1/2}\over\cosh r}, \quad {\rm and}\cr
N^{\phi}_{\pm}&={\sqrt2\over\cosh^2\!r},\cr}
$$
where~$N^a_+$ is tangent to the outgoing null geodesics and~$N^a_-$
to the ingoing ones (both vectors point in the past direction).
The expansion of the two systems may be computed to be
$$
\theta_{\pm}\equiv \nabla\!_aN^a_{\pm} =
\pm\left[{1-2\sinh^2\!r\over\sinh r\sqrt{1-\sinh^2\!r}}\right].
$$

Thus, the expansion of the outgoing null geodesics is positive for
small~$r$, becomes zero at~$r=r_0$ (where $\sinh^2\!r_0=1/2$), and
negative
thereafter. I.e., the past light cone of a point on the~$r=0$ axis
reconverges for $r>r_0$.
Meanwhile the ingoing null geodesics start off with negative expansion
for small~$r$. The expansion for this system, too, becomes zero
at $r=r_0$. Thus, at any time~$t$, the surface given by $r=r_0$
is a marginally trapped surface. It is interesting to note that though
the outgoing null geodesics are converging (i.e.,~$\theta<0$) for $r>r_0$,
$r$~continues to increase for these geodesics. And the ``focus'' that
the geodesics come to occurs on the surface given by $r=\log(1+\sqrt2)$.

\Sectionvar Appendix C: A Single Beginning?

Singularity theorems in general relativity often confine themselves
to proving the existence of one incomplete causal geodesic. In what
sense, then, can a cosmological singularity theorem be said to show
that the universe {\sl as a whole\/}~[\Linde] had a single singular
beginning?

Although it doesn't appear possible to conclusively prove the necessity
of a single beginning without resorting to model-dependent calculations,
there are pieces of evidence that suggest that the existence of
a single beginning is a plausible consequence of the singularity
theorems. For instance, Theorem~3 (and its extension in
section\(6)) shows that the past light cone of any point~$p$
satisfying assumption~D must be incomplete. Arguments given
elsewhere~[\Vilenkin\,\BVTwo] show that almost all points in the
inflating region of an inflationary spacetime will satisfy
assumption~D. It follows that each of these points must have an
initial singularity (i.e., a past-incomplete causal geodesic)
somewhere to the past. This does not prove that all these
singularities lie on one spacelike hypersurface, but it does make such
a scenario seem plausible.

Stronger evidence for this scenario may be obtained by adding to Theorem~4
the assumption that $\cal S$ is a global Cauchy surface and requiring that
$K$ be non-zero (it is not necessary then to assume that $\cal S$
is compact, making the result applicable to both open and closed universes).
It follows by a standard argument from the altered assumptions
that no timelike geodesic that emanates orthogonally from $\cal S$
can exist for a proper time greater than
$\tau_{\rm max}\equiv -(3/K)\coth^{-1}\!(\theta_{\rm max}/K)$ to the
past of~$\cal S$, where $\theta_{\rm max}$ is the largest value of the
divergence of
the geodesics at~$\cal S$ (i.e., it represents the least past-convergent
of these geodesics). For, suppose that there is a point~$r$ on one of these
geodesics a proper time $\tau>\tau_{\rm max}$ to the past of~$\cal S$.
Since $\cal S$ is a global Cauchy surface there must be a maximal
timelike curve~$\gamma$ between~$r$ and~$\cal S$~[\HE]. This curve
must have length greater than~$\tau_{\rm max}$ and it
must intersect $\cal S$ orthogonally~[\HE].
But then $\theta$ must diverge to $-\infty$
on $\gamma$, between $\cal S$ and~$r$. This means that~$\gamma$ cannot be
maximal~[\HE].

The result shows that the initial singularity for each
observer cannot lie arbitrarily far to the past. This is further
evidence that it seems reasonable to infer
that the classical universe did indeed have a single beginning.

\Sectionvar Acknowledgements

This paper arose out of discussions with Alex Vilenkin. It is a pleasure
to thank him for stimulating my interest in the questions discussed here,
for offering very many helpful comments and suggestions, and for reading
the manuscript. I also thank the Institute of Cosmology at Tufts
University for its warm hospitality over the period when this work
was done, and the High Energy Theory Group at Brookhaven National
Laboratory for its continued support.

\EndPaper